# Development of a High-Resolution, High-Dynamic-Range Charge Detector for Ion Beam Monitoring

O. Adriani[a,b], E. Berti[a], P. Betti[a], J. Casaus[d], R. D'Alessandro[a,b], S. Detti[a], C. Diaz[d], J. Marin[d], G. Martinez[d], N. Mori[a], L. Pacini[a], C. Pizzolotto[c], A. Tiberio[a,b], M. Scaringella[a], O. Starodubtsev[a,1], G. Zampa[c], N. Zampa[c]

[a] *Istituto Nazionale di Fisica Nucleare, Sezione di Firenze,*
 *I-50019 Sesto Fiorentino, Florence, Italy*

[b] *Department of Physics and Astronomy, University of Florence,*
 *I-50019 Sesto Fiorentino, Florence, Italy*

[c] *Istituto Nazionale di Fisica Nucleare, Sezione di Trieste,*
 *via A. Valerio 2, Trieste, Italy, I-34149 Trieste, Italy*

[d] *Centro de Investigaciones Energéticas, Medioambientales y Tecnológicas (CIEMAT),*
 *E-28040 Madrid, Spain*

*E-mail*: `starodubtsev@fi.infn.it`

ABSTRACT: We present an innovative charge detector with high resolution and wide dynamic range designed to fulfill the requirements of a monitoring system for a high energy ion beam. The detector prototype, constructed using Si photodiodes and a custom readout electronics, underwent extensive testing during HERD and AMS beam tests at CERN SPS facilities. Initial testing showcased the detector's exceptional performance, emphasizing both high resolution and a dynamic range capable of measuring nuclei with atomic numbers ranging from 1 to 80. The prototype's compatibility with fast, quasi real-time data analysis qualifies it as an ideal candidate for online applications. This article presents the results from the testing phase of the prototype, highlighting its capabilities and performance. Ongoing detector development, potential applications, and future developments aimed at enhancing the detector's functionality and versatility are also discussed.

---

[1] Corresponding author.

# Contents

## 1. Introduction

This work has been initiated within the HERD collaboration by the group of INFN section of Florence in cooperation with INFN section of Trieste and CIEMAT, Madrid, Spain. The High Energy cosmic-Radiation Detection (HERD) facility, proposed as a component of China's Space Station (CSS), is scheduled for operation starting around 2027 for a decade. The primary scientific objectives of HERD encompass groundbreaking research areas, including the indirect search for dark matter with unparalleled sensitivity, precise measurements of cosmic ray spectra and compositions up to the knee energy region, and gamma-ray monitoring alongside a comprehensive sky survey [1].

The HERD detector architecture is based on a homogeneous, deep, 3D segmented calorimeter, complemented with scintillating fiber trackers, anti-coincidence scintillators, silicon charge detectors, and a transition radiation detector [2]. Designed to maximize the geometric acceptance, HERD extends measurement capabilities to substantially higher energies with respect to the current generation of direct-measurement experiments, potentially reaching the PeV region for light nuclei. This ambitious endeavour aims to significantly advance our understanding of cosmic ray propagation and acceleration within the Galaxy, with the precise measurement of proton and nuclei fluxes surpassing hundreds of TeV per nucleon. Furthermore, exploring electron flux in the multi-TeV range provides a tool for discerning signatures of dark matter and nearby astrophysical sources. Additionally, the wide field of view of HERD allows a comprehensive monitoring of the gamma-ray sky across energy ranges spanning from a few hundred MeV up to 1 TeV.

Throughout the research and development phase, various system components and subdetectors underwent extensive testing at different beam facilities at CERN Proton Synchrotron (PS), Super Proton Synchrotron (SPS) [3], and other facilities [4, 5]. Notably, tests with ion beams at the SPS facility highlighted a critical need for precise beam quality monitoring. Although the HERD apparatus incorporates a charge detector, its dynamic range is currently aiming to reach Z~26, and its reliance on silicon microstrip sensors implies complex calibration procedures and time-consuming data analysis [6]. These constraints make the HERD charge detector unsuitable for real-time beam monitoring applications. To address this challenge, we have developed a real-time, easy to use and low-cost charge tagger solution.

## 2. Detector design

The charge tagger exploits the phenomenon of direct ionization within the depleted region of silicon photodiodes. In constructing the detector, we made use of the photodiodes originally employed in the CaloCube R&D project [7], which validated the 3-D calorimetric approach implemented in HERD. The initial prototype comprised six blinded photodiodes (PDs) arranged in a one-dimensional array (Fig. 1). Our design prioritized simplicity and lightweight construction, achieved through a homemade mechanical structure that facilitates an easy mounting and dismounting. Furthermore, the structure was engineered with a relatively small material budget to minimize the nuclei fragmentation within the detector.

To ensure seamless integration, we opted to utilize the same front-end electronics deployed in the HERD calorimeter PD subsystem [8]. This strategic choice enabled us to capitalize on existing infrastructure, resulting in a high dynamic range, low noise operation. Additionally, compatibility with the overall trigger and data acquisition system was effortlessly achieved. This streamlined integration underscores the charge tagger's potential for practical deployment within the HERD instrument suite.

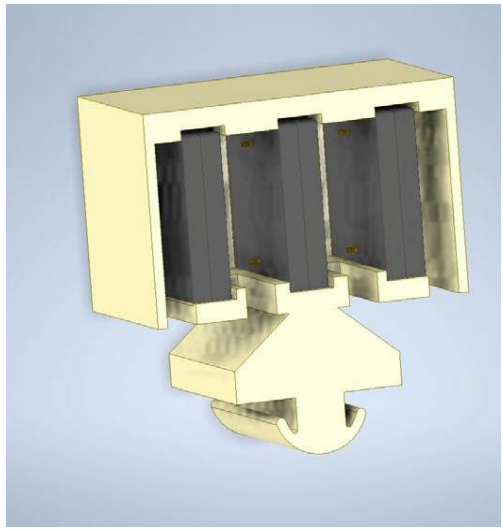

**Figure 1.** First prototype mechanical design. 6 blinded PDs in their package (the black elements in the figure) are mounted in 3 pairs.

### 2.1 Diode

The VTH2090 photodiode by Excelitas is a PIN photodiode featuring a 9.2×9.2 mm$^2$ active area, housed within a black ceramic package with an epoxy window [9]. Although detailed information regarding the diode thickness is unavailable, we conducted a series of capacitance measurements to infer its characteristics.

Typically, the depletion voltage is determined by analysing the inverse squared capacitance as a function of the applied voltage and fitting two lines to find their intersection point [10]. This method allows us to extract the depletion thickness using the equation:

$$\omega = \sqrt{\frac{2\varepsilon}{e|N_{eff}|}V} \quad (1)$$

where:

    ω is the depletion width,
    $\epsilon$ is the permittivity of the material,
    e is the elementary charge,
    $N_{eff}$ is the effective doping concentration, and
    V is the applied voltage.

Results from the capacitance measurements and subsequent calculations revealed a depletion depth of approximately 220 μm (Fig. 2) and a capacitance of around 47 pF. The slight rise in capacitance after full depletion is attributed to edge effects. Although a capacitance plateau could, in principle, be achieved at higher bias voltages, this was not possible in our case due to the onset of the breakdown effect. The plotted curve shows the initial signs of breakdown as a slight decrease in capacitance around a bias voltage of 150 V. Additionally, considering the thickness of the ceramic package of about 1 mm and the total package thickness of 2 mm (including the diode and epoxy window) provides a complete understanding of the photodiode's physical dimensions. These measurements and calculations provide valuable insights into the operational

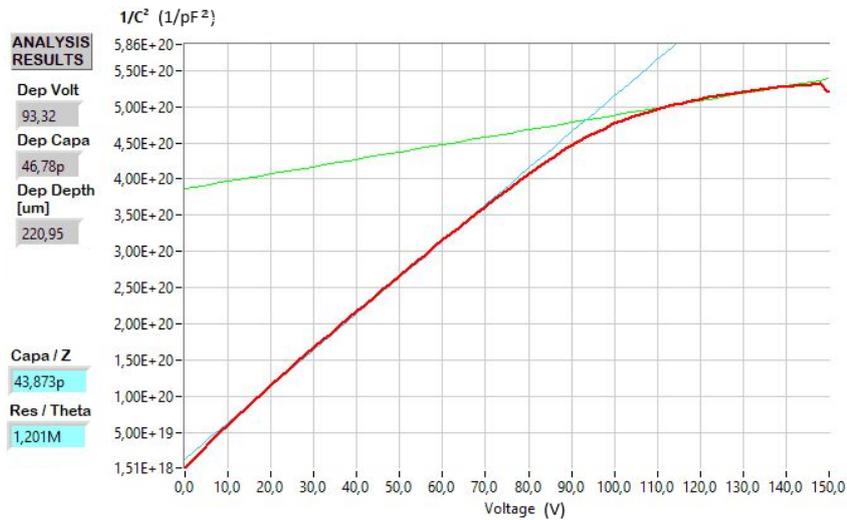

**Figure 2.** Inverse squared capacitance vs. voltage measurement results with depletion voltage fit and depletion depth calculation.

characteristics of the VTH2090 photodiode.

## 2.2 Electronics

The charge tagger was connected to the front-end electronics using the same flat kapton cables used for the readout of the HERD calorimeter.

The front-end chip utilized was the HiDRA version 2 specifically designed for space applications [11]. The HiDRA chip is an advanced evolution of the CASIS (CAlorimetry in SIlicon for the Space) chip [12], which is an Application Specific Integrated Circuit (ASIC) specifically designed for space calorimetry by the INFN section of Trieste. This chip boasts several key features:
- Double gain Charge Sensitive Amplifier (CSA) with automatic-gain selection circuitry.
- High dynamic range from few fC to 52 pC.
- Low power consumption (~3.75 mW/channel)
- Low noise, with an equivalent noise charge (ENC) of approximately 2500 e-
- 16 input channels
- Self-trigger system
- Maximal acquisition rate 1000 Hz

Fig. 3 left illustrates the block scheme of a single channel of a HiDRA 2 ASIC, while Fig. 3 right shows one front-end board containing four chips. The selection of this electronic component was based on its compatibility and performance with the VTH20290 PDs.

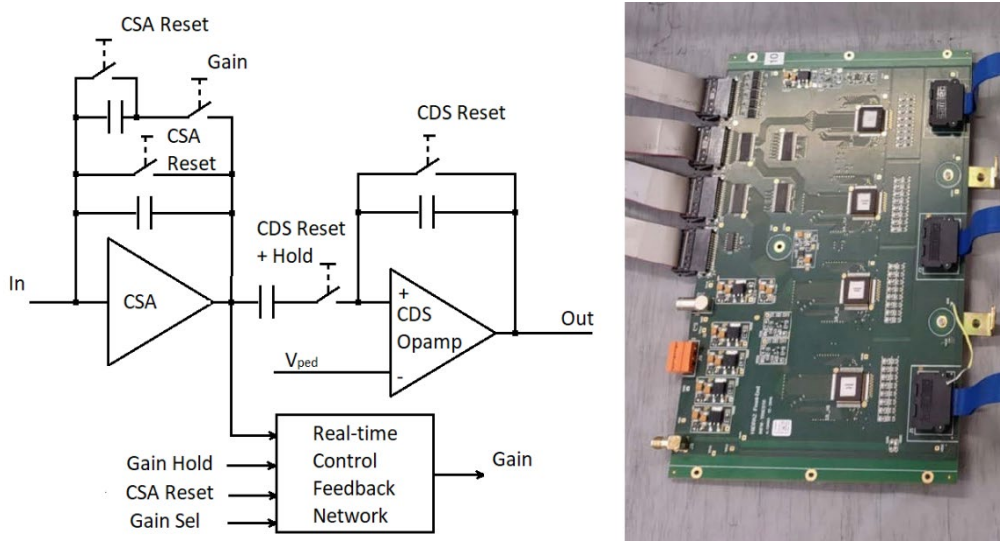

**Figure 3.** Left: the block scheme of a single channel of a HiDRA 2. Right: HERD calorimeter front-end board containing four chips.

The chip consists of a charge sensitive amplifier (CSA) with an automatic double-gain selection: when the CSA output signal exceeds a given threshold, the gain is reduced by a factor of 20. Laboratory tests have shown that the uncertainty of this value is below 1% [8]. The CSA output is then connected to a correlated double sampling circuit (CDS), which samples the signal within an integration window of a few microseconds, and to a fast trigger circuit. The latter comprises a common threshold generator and, for each channel, a processing chain made up of a signal amplifier and a comparator. One HiDRA-2 chip has 16 input channels, with their analog output multiplexed to a single output pin.

The trigger circuit includes a common threshold generator and, for each channel, a processing chain consisting of a signal amplifier and a comparator. To decrease the power consumption and reduce the number of chip output pins, the outputs of two adjacent comparators are ORed together and then converted to differential CMOS signals for further external processing.

Four trigger threshold values are available, corresponding to the internally generated threshold voltage multiplied by 1, 1.5, 2, and 2.5. These values can be selected by appropriately setting the corresponding digital inputs on the chip.

The architecture used results in very low noise, an excellent signal-to-noise ratio, and a high dynamic range, which is at least a factor of 5 higher than commercially available solutions such as the SKIROC 2A [13] or VA32HDR14.3 [14] ASICs.

## 3. Prototypes

Two prototypes were built. The first one was constructed following the design described above. For the second prototype, we maintained consistency by utilizing the same mechanical structure and photodiode model as those employed in the initial design but with selected diodes as explained below.

The first prototype was operated with a bias voltage of 40V, optimized for calorimetric light detection, well below the full depletion voltage of approximately 90V. For light detection, achieving full depletion is not an absolute necessity as in light detection mode they typically require very low bias or no bias at all. We deliberately chose to operate at this lower voltage level to stay comfortably below the maximum operating voltage limit specified in the datasheet while ensuring a reasonable diode capacitance to maintain a low noise performance.

However, in the context of charge measurements, full depletion becomes critical as the charge release depends on depletion thickness. Full depletion ensures a larger and more uniform signal, thereby enhancing measurement accuracy. Using these considerations, the operating bias voltage of 100 V was applied for the second prototype. Consequently, for the second prototype, all PDs were carefully selected to feature a breakdown voltage higher than 100 V to ensure full depletion.

## 4. Data analysis routine

For the first prototype, a relatively simple and efficient analysis routine was employed:
- Signal amplitude evaluation in ADC for each of the six blinded PDs, ensuring it exceeded three times the noise threshold;
- Signal conversion from ADC to Minimum Ionizing Particle (MIP) units using prior calibration. The conversion parameters from ADC to MIP were obtained for each diode by analysing the carbon peak in the fragmented SPS beam;
- Conversion of the signal from MIP to atomic number (Z);
- A minimum of four consistent PD with the signals above the noise are required for analysis. The consistency requirement is defined as a signal difference of less than 1.5 charge units;

This straightforward data analysis approach facilitated rapid processing and interpretation of detector signals, ensuring reliable results for subsequent analysis and interpretation.

We introduced several enhancements to the analysis routine for the second prototype. Specifically, we applied all the previously described selection criteria while increasing the requirement for the number of consistent PDs to six. This means that all diodes used in the detector must produce consistent signals, see Fig.1. Only events meeting these criteria were selected. While the efficiency of this selection method is currently less than 50%, we are actively working on optimizing the routine.

This refinement in the analysis routine aims to improve the robustness and reliability of the data obtained from the charge detector, ensuring more accurate and consistent results for subsequent analysis and interpretation.

## 5. Test results

The first prototype underwent testing in various experiments at the SPS facility using ion beams:

- In 2022, the HERD experiment utilized the prototype as a nuclei tag to assess the non-linearity of the LYSO scintillator, the active material in the HERD calorimeter.
- In 2023, the HERD experiment employed the prototype to evaluate the performance of a large-scale HERD prototype with nuclei.

During these tests, the ion beam was generated by colliding a lead primary beam of 150 GeV/n with a 4 cm Beryllium target. The resulting secondary fragmented beam was selectively filtered based on rigidity using magnetic optics, enabling a precise selection of the mass-to-charge ratio, A/Z, of the fragmented products. The fragmented beam consisted of a wide range of elements with Z values spanning from 1 to 82 [15]. Different values of A/Z have been used to tune the composition of the fragmented beam.

The second prototype was utilized outside the HERD experiment. In 2023, thanks to the excellent performance demonstrated during the 2022 HERD beam test, we received a proposal from the AMS-02 project to participate in ion beam tests of silicon microstrip detectors and provide independent beam monitoring measurements. The data obtained facilitated an independent evaluation of charge tagging and charge reconstruction efficiency of the AMS silicon microstrip detectors.

For the determination of the particle charge for each event, various approaches were explored, with the truncated mean charge value emerging as the most effective. The following figures show the results obtained from the first prototype during the initial 2022 HERD beam test at the SPS ion beam.

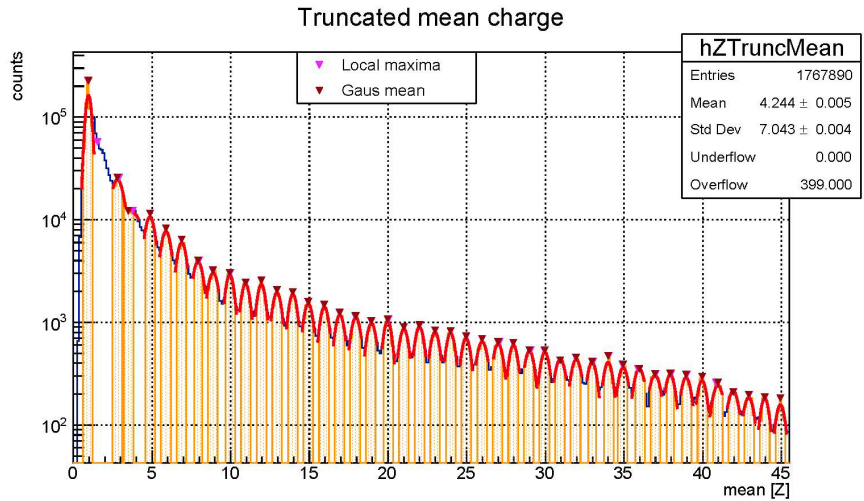

**Figure 4.** First prototype. Truncated mean charge distribution measured for an ion beam with A/Z ~ 2.2

Fig. 4 illustrates the truncated mean charge distribution measured for an ion beam with an A/Z ratio of approximately 2.2. The measured beam composition with nuclei peaks (blue histogram) ranges from Z=1 up to Z=45. A Gaussian fit was applied to each peak (red curves, with the yellow areas representing the 1-sigma confidence level region; no yellow area is present for failed/bad fits). This analysis routine allows for the precise determination of the Z value of the nuclei.

During this test, the detector was positioned downstream of several other detectors, resulting in a significantly fragmented and contaminated beam. As a consequence, achieving a reliable fit for the low Z peaks was not feasible.

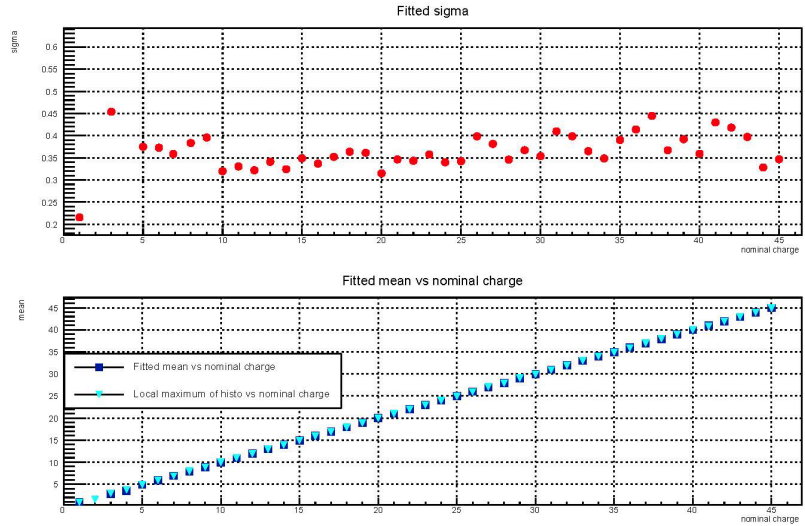

**Figure 5.** First prototype. Measured resolution (upper image) and linearity (lower image) for an ion beam with A/Z ~ 2.2

Fig. 5 demonstrates the Z resolution (upper image) and charge linearity (lower image) of the first prototype. The first version exhibited an almost-flat charge resolution better than 0.5 ΔZ up to Z=60. The linearity was analysed in two ways: by using the local maximum of the measured peaks and by using the mean value of the Gaussian fit. The prototype shows excellent linearity across the entire range of measured charges for both methods.

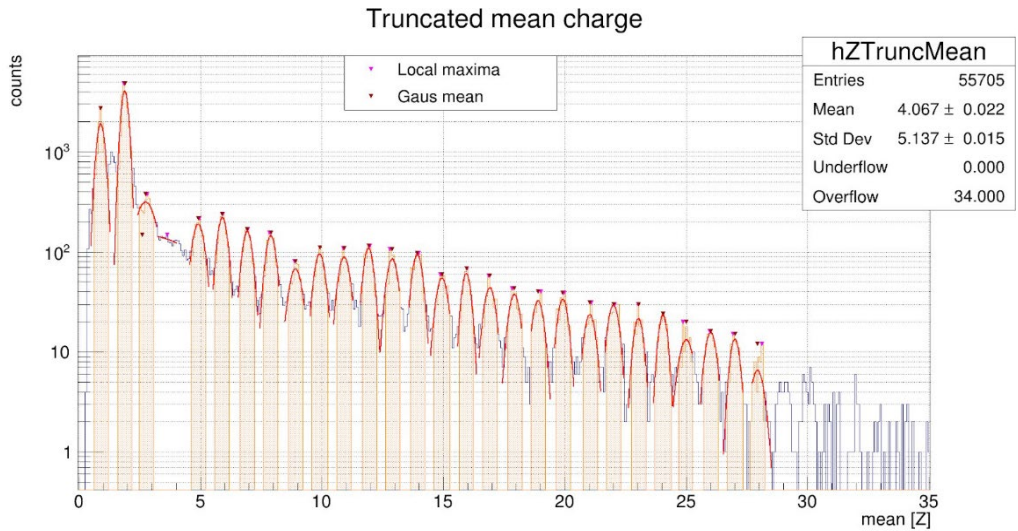

**Figure 6.** Second prototype. Truncated mean charge distribution for an ion beam with A/Z ~ 2

The second prototype was tested with ion beams characterized by A/Z values of approximately 2 and 2.2. Figures 6 and 7 depict the truncated mean charge values for both beams, following the same plotting logic used for the first prototype. Figure 8 illustrates the charge resolution and linearity for the A/Z~2.2 beam, showcasing significant improvement with a resolution better than 0.3 ΔZ from Z=5 to Z=75. This enhancement is attributed to the higher bias voltage and refined event selection strategies implemented in the second prototype. The refined analysis routine described in section 4 was used for the second prototype and aims to enhance the robustness and reliability of the data obtained from the charge detector, ensuring more accurate and consistent results for subsequent analysis and interpretation.

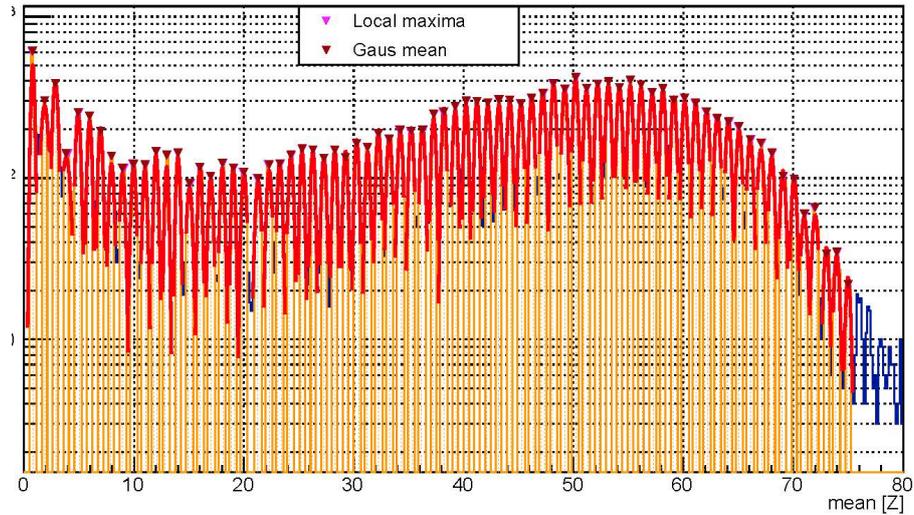

**Figure 7.** Second prototype. Truncated mean charge distribution for an ion beam with A/Z ~ 2.2

In this preliminary analysis, the resolution of nuclei below Z=5 is not accurately evaluated, as the current fitting procedure does not account for asymmetric distributions and a background subtraction procedure is not implemented. Despite the analysis not being optimized in this region, the low Z nuclei peaks are clearly identified by the detector.

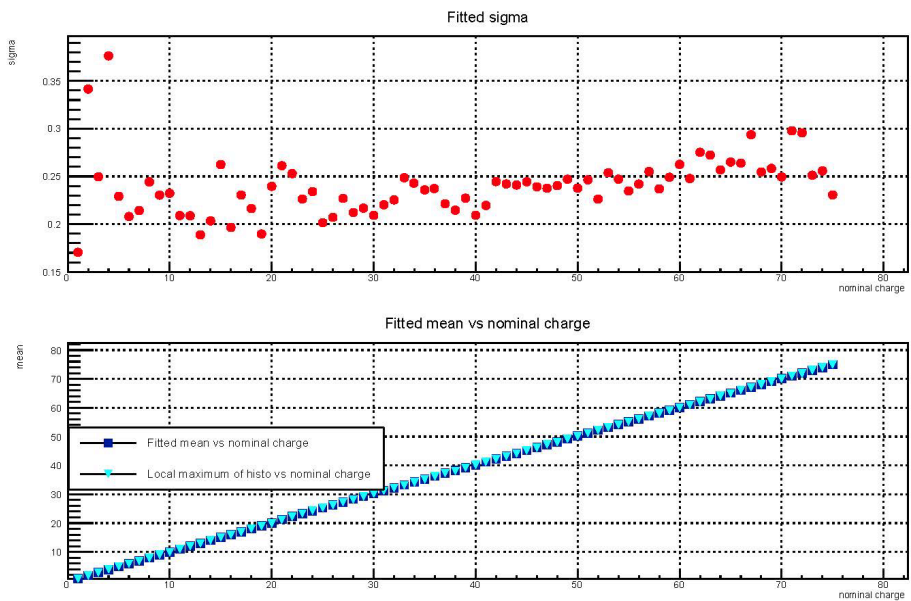

**Figure 8.** Second prototype. Measured resolution and linearity for an ion beam with A/Z ~ 2.2.

## 6. Further development

In preparation for future applications, we work mainly on two items: sensors and readout electronics.

### 6.1 Optimized Sensor Configuration

We are currently focusing on redesigning the sensor configuration to enhance performance and versatility. Our new design incorporates a series of 3×3 matrices, as depicted in Fig. 9. To minimize the material budget along the beam line and mitigate nuclei fragmentation, we intend to utilize diodes without packaging. Additionally, we plan to optimize the size of the diodes, considering options such as 5×5 or 10×10 mm$^2$ configurations. Adopting a modular design approach will facilitate easy customization, enabling us to adjust the number of matrices and the quantity and size of diodes as needed for specific applications. This redesign aims to maximize detection efficiency and improve overall detector performance.

### 6.2 Readout Electronics Upgrade

Recognizing the limitations of the current electronics, we are actively working to optimize it. The current system has several pros and cons. The advantages are a very high dynamic range and low power consumption, while the disadvantages are a limited acquisition rate and limited acquisition efficiency.

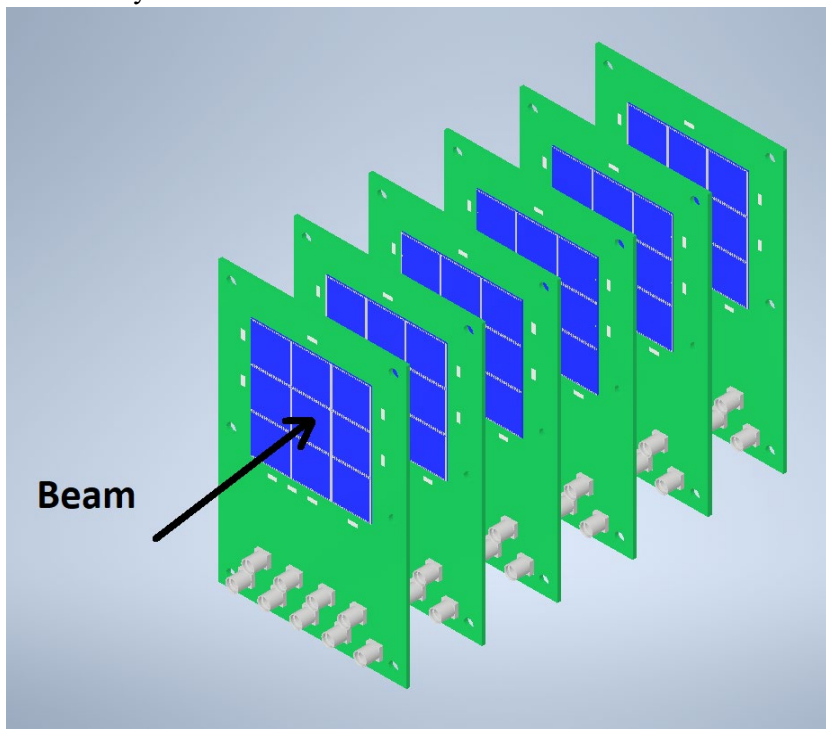

**Figure 9.** New matrix design example.

The acquisition efficiency is limited because approximately 10% of the events cannot be used for analysis. This inefficiency arises because the Charge Sensitive Amplifier (CSA) must be periodically reset. Consequently, during asynchronous triggers, a small number of events cannot be properly integrated. An in-depth discussion of this issue goes beyond the scope of this article, but details about the chip features and its application to particle acquisition can be found in [16].

Therefore, our current focus is on optimizing the readout electronics to better meet the more demanding requirements of potential future applications, such as beam monitoring at accelerators. The specific requirements for optimization will depend on the application, and we are considering two main directions for the most likely scenarios:
- High Acquisition Rate: Enhancing the system to handle a higher rate of data acquisition.
- Extended Dynamic Range: Achieving this by using a second chip with lower gain connected to the same diode.

These modifications are expected to significantly improve the usability and performance of the readout electronics for beam monitoring applications.

## 7. Conclusions

The initial prototypes of our charge detector have yielded promising results, demonstrating high performance across several key metrics. The design of the detector has proven effective, providing a very high dynamic range, good Z resolution, and excellent linearity. Notably, the simplicity and efficiency of our data analysis process differentiates our detector from silicon microstrip detectors. Our detector offers significant advantages in terms of simplicity in both application and data analysis. This is primarily attributed to the absence of a charge-sharing mechanism, which eliminates the need for readout element equalization or a series of complex corrections such as position, angular, velocity, rigidity, and time-dependent corrections [6, 17].

Our detector showcases a higher dynamic range, accommodating Z values up to 80 compared to the typical Z range of 28-30 for strip sensors read out by commonly available commercial frontend chips. The charge distribution obtained by one of the HERD silicon strip detectors prototypes developed by INFN section of Perugia [15] ranges from Z=1 up to Z=10, further highlighting the effectiveness of our charge tagger. Our detector demonstrates a linear Z resolution of approximately 0.25 $\Delta Z$ across the entire Z range, in contrast to the non-linear charge resolution obtained with HERD Si strip charge detector [6], where it is under 0.2 $\Delta Z$ for carbon and escalates for higher Z elements.

However, despite these successes, we have identified several points that can be improved, with the most significant being related to readout electronics. While the HERD calo electronics were initially selected for their high dynamic range and integration with HERD DAQ, they were primarily designed for space applications and are not optimized for accelerator beam applications. While the current ASIC can easily cope with the relatively low rate of high-energy (more than tens of GeV) particles in space, it falls short when used for monitoring particle beams from accelerators, having a maximal safe rate of acquisition of about 1 kHz. Furthermore, the chip operation mode exhibits temporal patterns, compromising data integrity as events occurring at the beginning and end of the asynchronous charge integration interval may experience incomplete charge collection, which effectively results in the exclusion of approximately 10% of acquired data.

With the improved sensor design, and after properly addressing the limits of the readout electronics described above, we see several potential application fields:
- Online monitoring of ion beam composition (e.g. for the fragmented beam in the North Area experimental facility at CERN).
- Charge tagger for users at North Area experimental facility of CERN: Providing accurate nuclei tagging on an event-by-event basis.

- Online multiplicity monitor for various beams, for example, the high-multiplicity electron-positron beams at BTF [18].
- Space-based and balloon-borne cosmic ray studies.

The proposed detector concept is modular and scalable. Coupled with the low cost and wide variety of available sensors, it offers numerous potential applications. The readout electronics should be optimized for each specific case to ensure optimal performance.

**References**

[1] A. E. Berti, N. Mori, L. Pacini and O. Starodubtsev, *The High Energy Cosmic-Radiation Detection HERD facility: a future space instrument for cosmic-ray detection and gamma-ray astronomy*, in proceedings of *27th European Cosmic Ray Symposium - ECRS* July 25-29, 2022, Nijmegen, the Netherlands. PoS(ECRS)146 .

[2] N. Mori, L. Pacini on behalf of the HERD collaboration, *The High Energy Cosmic-Radiation Detection (HERD) facility for direct cosmic-ray measurements,* in proceedings of *41st International Conference on High Energy physics - CHEP2022*, July 6-13 2022 Bologna, Italy PoS(ICHEP2022)123 .

[3] https://home.cern/news/news/accelerators/accelerator-report-getting-lead-ions-ready-physics

[4] P. Betti et al., *Photodiode Read-Out System for the Calorimeter of the Herd Experiment*, Instruments 2022, 6(3), **33** [instruments6030033].

[5] Y. W. Dong, Z. Quan, J. J. Wang, M. Xu, S. Albergo, F. Ambroglini, et al., *Experimental verification of the HERD prototype at CERN SPS,* Space Telescopes and Instrumentation 2016, UK: Ultraviolet to Gamma Ray, Edinburgh, 2016. DOI:10.1117/12.2231804.

[6] Wei-Shuai Zhang et al., *A novel charge reconstruction algorithm applied to the HERD prototype silicon charge detector.* 2024 NIM A Volume 1064, 2024, 169346. DOI:10.1016/j.nima.2024.169346.

[7] O. Adriani et al., *The CALOCUBE project for a space based cosmic ray experiment: design, construction, and first performance of a high granularity calorimeter prototype* 2019 JINST 14 P11004 DOI:10.1088/1748-0221/14/11/P11004.

[8] O. Adriani et al., *Development of the photo-diode subsystem for the HERD calorimeter double-readout,* 2022 JINST **17** P09002. DOI:10.1088/1748-0221/17/09/P09002.

[9] https://www.datasheetcatalog.com/datasheets_pdf/V/T/H/2/VTH2090.shtml

[10] https://cds.cern.ch/record/2318208/files/MikeJonesMSc_2.pdf pp.23-24

[11] ] L. Pacini et al., *Design and expected performances of the large acceptance calorimeter for the HERD space mission*, PoS ICRC2021 (2021) 066.

[12] V. Bonvicini, G. Orzan, G. Zampa and N. Zampa, *A double-gain, large dynamic range front-end ASIC with A/D conversion for silicon detectors read-out*, IEEE Trans. Nucl. Sci. 57 (2010) 2963.

[13] https://www.weeroc.com/read_out_chips/skiroc-2a/

[14] https://ideas.no/products/va32hdr14-3/

[15] A. Oliva et al., *The silicon charge detector of the high energy cosmic radiation detection facility*. Proceedings of 38th International Cosmic Ray Conference - ICRC2023, Nagoya, Japan, PoS(ICRC2023)26.

[16] O. Adriani et al., *The CaloCube calorimeter for high-energy cosmic-ray measurements in space: performance of a large-scale prototype*, 2021 JINST 16 P10024 [arXiv:2110.01561].

[17] B. Alpat et al., *Charge determination of nuclei with the AMS-02 silicon tracker* 2005 NIM A Volume 540 Issue 1, Pages 121-130, DOI:10.1016/j.nima.2004.11.012.

[18] https://btf.lnf.infn.it/